\documentclass[prd,aps,preprint,floats,tightenlines]{revtex4}

\usepackage{epsfig}

\begin{document}
 
\pagestyle{empty}

\preprint{
\noindent
\begin{minipage}[t]{3in}
\begin{flushright}
LBNL--61605 \\
hep-ph/0609133 \\
\end{flushright}
\end{minipage}
}

\title{Partial waves of baryon-antibaryon in three-body $B$ meson decay}

\author{
Mahiko Suzuki
}
\affiliation{
Department of Physics and Lawrence Berkeley National Laboratory\\
University of California, Berkeley, California 94720
}


\date{\today}

\begin{abstract}

The conspicuous threshold enhancement has been observed in the 
baryon-antibaryon subchannels of many three-body $B$ decay modes. 
By examining the partial waves of baryon-antibaryon, we first 
show for $B^{\pm}\to p\overline{p}K^{\pm}$ that the $pK^{\pm}$ 
angular correlation rules out dominance of a single $p\overline{p}$ 
partial wave for the $p\overline{p}$ enhancement, for instance, the 
resonance hypothesis or the strong final-state interaction in 
a single channel. The measured $pK^{\pm}$ angular correlation turns
out to be opposite to the theoretical expectation of a simple 
short-distance picture. We study the origin of this reversed angular 
correlation in the context of the $p\overline{p}$ partial waves and 
argue that $N\overline{N}$ bound states may be the cause of this sign 
reversal. Dependence of the angular correlation on the $p\overline{p}$ 
invariant mass is important to probe the underlying issue from 
the experimental side.   
\end{abstract}
\maketitle

\pagestyle{plain}

\setcounter{footnote}{0}

\section{Introduction}
            
 In the baryonic $B$ decay the three-body modes dominate over the 
two-body modes. Furthermore, in the three-body decay, the 
baryon-antibaryon pair is copiously produced at small invariant mass 
near the threshold\cite{Belle1,Belle2,BaBar,Belle3,Lee}. Various theoretical 
ideas\cite{Soni,Glueball,TH,pole,Chengsum}, some kinematical and others 
dynamical, were proposed for this threshold enhancement. A simple 
short-distance (SD) argument can explain qualitatively both the dominance 
of three-body modes and the threshold enhancement of baryon-antibaryon:
To produce a baryon and an antibaryon in the two-body decay (Fig. 1a), 
one energetic $q\overline{q}$ pair must be emitted back to back by 
a gluon so that the gluon emitting the $q\overline{q}$ pair is highly 
off mass shell. The hard off-shell gluon suppresses two-body decay 
amplitudes by the power of $\alpha_s/t$, where $t$ is the four-momentum 
square transferred through the gluon. In the three-body decay with an
additional meson (Fig. 1b), a baryon-antibaryon pair can be emitted 
collinearly against the energetic boson in the final state. In this 
configuration a quark and an antiquark are emitted by a gluon nearly 
in the same direction so that the gluon is close to the mass shell 
and the short-distance suppression does not occur. Consequently the
$p\overline{p}$ of small invariant mass is strongly favoured.

\noindent
\begin{figure}[h]
\hskip 3cm
\epsfig{file=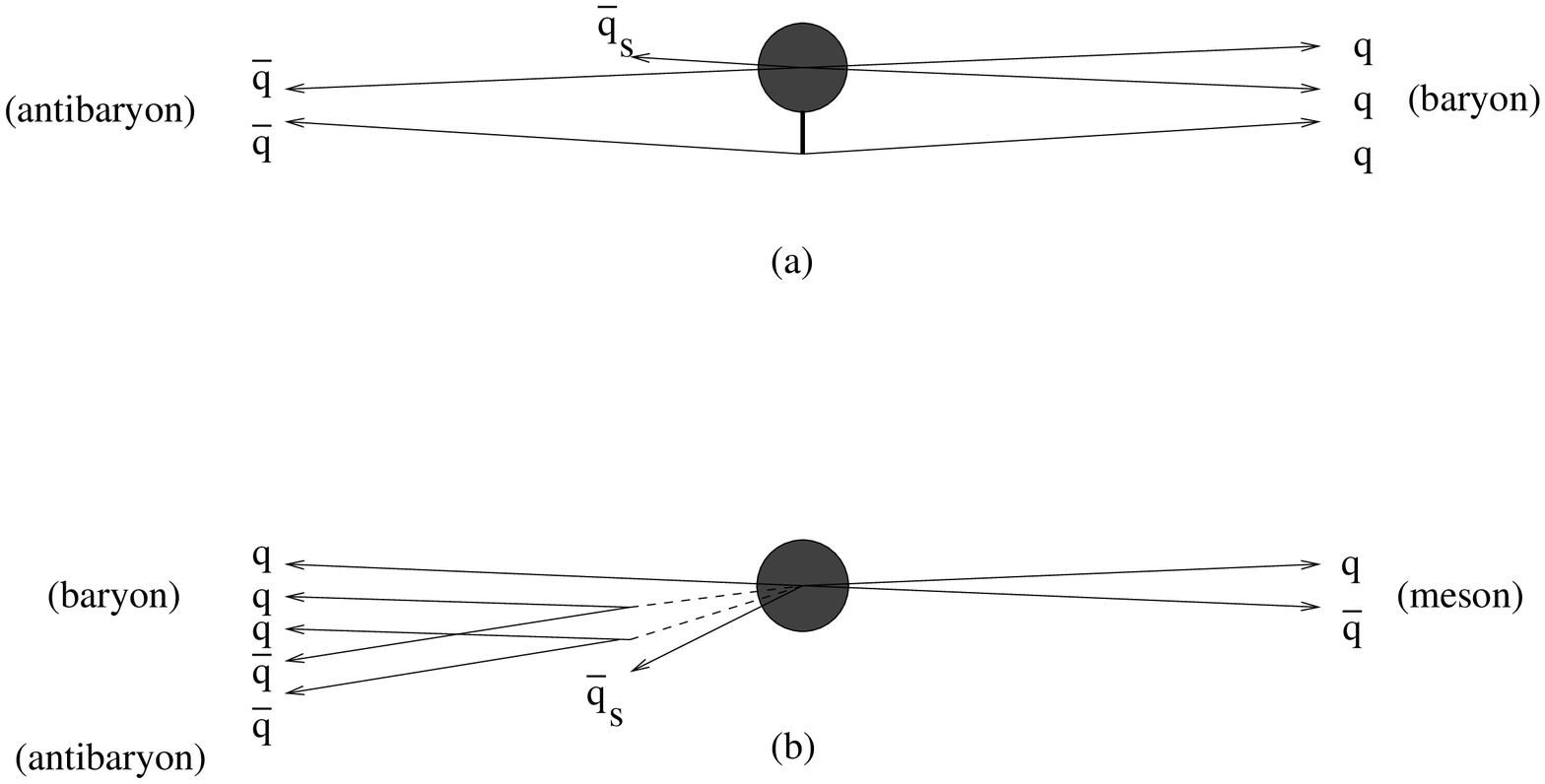,width=8cm,height=5cm}
\caption{Short-distance picture in quarks and antiquarks for
(a) two-body baryonic decay and (b) three-body baryonic decay.  
In the two-body decay (a) the fat virtual gluon (the thick 
vertical solid line) must split into $q\overline{q}$ while 
in the three-body decay (b) nearly on-shell gluons (broken lines) 
turn into $q\overline{q}$. The slow spectator antiquark is denoted by 
the short line $\overline{q}_s$.
\label{fig:1}} 
\end{figure}
 
In addition to threshold enhancement, the angular correlation was 
measured between the final proton or antiproton and the boson in some 
modes, most clearly in $B^{\pm}\to p\overline{p}K^{\pm}$\cite{PL}. 
Then an intriguing puzzle\cite{Chengsum} has emerged in the preceding 
SD picture: In that picture, the antiproton momentum should point 
more likely to the direction of the $K^-$ momentum in the 
$p\overline{p}$ rest frame of $B^-(b\overline{u})$ decay. That is, the 
proton should tend to move away from $K^-$ in this frame. The reason 
is that the antiproton $\overline{p}$ picks up the slow
spectator $\overline{u}$-quark and therefore its momentum is smaller
on average than that of the proton $p$ in the $B^-$ rest frame. By
boosting the $B^-$ rest frame to the $p\overline{p}$ rest frame, we
reach this conclusion.

However, the Belle Collaboration showed exactly the opposite\cite{PL}; 
it is the proton that is emitted along $K^-$ in the $p\overline{p}$ 
rest frame. Belle selected the threshold events by making a cut in 
the $p\overline{p}$ invariance mass $m_{p\overline{p}} (< 2.85$ GeV), 
but did not give the angular correlation as a function of 
$m_{p\overline{p}}$ for the selected events.  Meanwhile BaBar gave 
a Dalitz plot of $p\overline{p}K^-$\cite{BaBar} from which one can 
read the same trend as Belle's angular dependence. 

Rosner\cite{TH} argued qualitatively in terms of quark diagrams\cite{Harari} 
and predicted this angular correlation with baryon production through 
diquarks. But the argument does not seem to work for all baryonic modes  
in its simple form. Cheng and others\cite{Chengsum} computed the decay 
amplitudes in the pole model with factorization, leaving out inelastic 
form factor terms\cite{pole,Chengsum}. Their result does not lead 
to the correct angular correlation in the case of $B^-\to p\overline{p}K^-$. 
The simple SD picture presented at the beginning is successful in the
angular correlation of most three-body baryonic modes, {\em e.g.,} 
$B^-\to\Lambda\overline{p}\gamma$, but fails notably for 
$B^-\to p\overline{p}K^-$. Its failure suggests us importance of 
long-distance (LD) effects somewhere in the decay process. Indeed, 
the fragmentation by quark diagram and the pole model both contain 
some of the LD effects in very different ways. In this paper we take 
a close look at this angular correlation of $B^{\pm}\to p\overline{p}K^{\pm}$ 
from the viewpoint of partial waves in general and try to resurrect the 
simple SD picture by incorporating an appropriate LD effect in it. 
  
  In our proposed analysis we first examine the partial-wave content 
of $p\overline{p}$ in $B^-\to p\overline{p}K^-$ (and its conjugate) 
and conclude purely kinematically that the $p\overline{p}$ 
enhancement cannot be a broad resonance. For the same reason we rule 
out strong final-state interaction (FSI) in a single $p\overline{p}$ 
partial wave as a cause of the enhancement. We shall observe that 
reversal of the angular correlation occurs if some LD effect flips 
relative signs of partial-wave decay amplitudes. Such sign flip may 
indeed occur if $N\overline{N}$ bound states exist in right channels. 
The recently discovered state $X(1835)$\cite{Ablikim} is a good candidate 
that may be responsible for the sign flip. If $X(1835)$ should be 
an $N\overline{N}$ bound state, we expect a similar bound state in 
other channels from our reasoning of the sign flip. 

\section{$p\overline{p}$ partial waves in $B^-\to\ p\overline{p}K^-$}

We study the angular correlation between the proton momentum 
and the kaon momentum in the rest frame of $p\overline{p}$ by 
choosing the $z$-axis along the $K^-$ momentum. 
(Fig. 2) 

\noindent
\begin{figure}[h]
\hskip 3cm
\epsfig{file=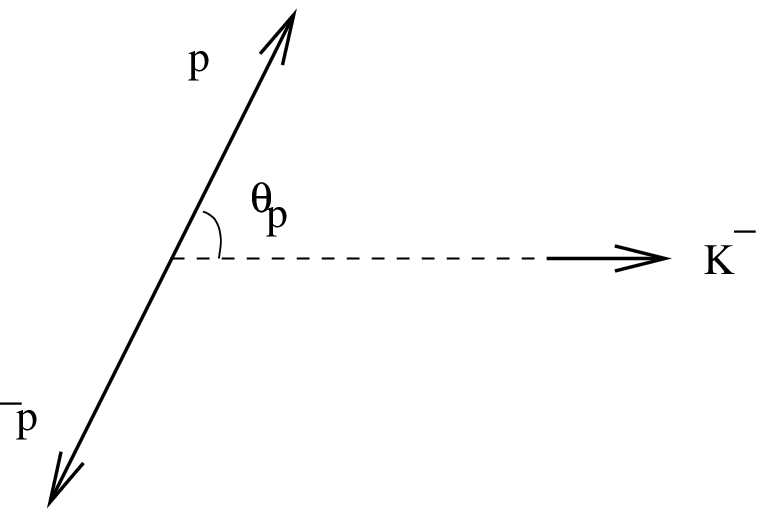,width=6cm,height=4cm}
\caption{The $pK^-$ angular correlation 
in the $p\overline{p}$ rest frame.
\label{fig:2}} 
\end{figure} 

 Since $B^-$ meson and $K^-$ meson are spinless and their momenta 
are both along the $z$-direction in the $p\overline{p}$ rest frame, the 
$z$-component of total angular momentum is zero for $p\overline{p}$ 
by $\Delta J_z = 0$ in this frame. Following the standard helicity 
formalism\cite{JW}, we can describe the angular dependence of the 
helicity decay amplitudes with Wigner's $d$-functions \cite{Rose} as 
\begin{equation}
A(B^-\to p\overline{p}K^-) = \sum_J
        A^{JK^-}_{\lambda_p\lambda_{\overline{p}};0}
        d^J_{0\lambda}(\theta_p)e^{-i\lambda\phi_p},\;\;\;
  (\lambda = \lambda_p-\lambda_{\overline{p}}),   \label{angular1}
\end{equation}
where $\lambda_p$ and $\lambda_{\overline{p}}$ are the helicities of 
$p$ and $\overline{p}$ in the $p\overline{p}$ rest frame, $(\theta_p, 
\phi_p)$ are the angles of the proton momentum in this frame (Fig. 2), 
and $A^{JK^-}_{\lambda_p\lambda_{\overline{p}};0}$ is a function 
of the $p\overline{p}$ invariant mass $m_{p\overline{p}}$. 
Since experiment does not measure helicity of proton nor antiproton
but sums over all helicity states in what follows, the differential 
decay rates are $\phi_p$ independent. Therefore
we need not specify the direction of $\phi_p=0$ in our case.\footnote{
We would have to fix the $\phi_p=0$ direction if a final particle spin 
is measured or if a final particle having spin undergoes a cascade 
decay and this decay angular correlation is measured.} 
Squaring the amplitude and 
summing over the $p\overline{p}$ helicities $\lambda_p$ and 
$\lambda_{\overline{p}}$, we obtain the differential decay rate:
\begin{equation}
  \frac{d\Gamma(\theta_p)}{dm_{p\overline{p}}d\Omega_p}\biggr|_{B^-
      \to p\overline{p}K^-}
   =\Gamma_0 \sum_{\lambda_p\lambda_{\overline{p}}}
 \Bigl|\sum_J A^{JK^-}_{\lambda_p\lambda_{\overline{p}};0}
 d^J_{0\lambda}(\theta_p)\Bigr|^2,         \label{angular2}
\end{equation}
where $\Gamma_0$ includes kinematical factors that depend on 
$m_{p\overline{p}}$. If we make the usual assumption that the strong penguin
interaction dominates in the decay $\overline{B}\to p\overline{p}
\overline{K}$, the CP-violating phases drop out of the decay rate. 
Under parity reflection the angle $\theta_p$ remains unchanged, 
while under charge conjugation the angle $\theta_p$ turns into 
$\pi-\theta_p$ of $B^+\to p\overline{p}K^+$ because of the interchange 
$p\leftrightarrow\overline{p}$ and $K^-\leftrightarrow K^+$. Therefore,
\begin{equation}
 \frac{d\Gamma(\theta_p)}{dm_{p\overline{p}}d\Omega_p}
      \biggl|_{B^+\to p\overline{p}K^+}=        
 \frac{d\Gamma(\pi-\theta_p)}{dm_{p\overline{p}}d\Omega_p}
      \biggl|_{B^-\to p\overline{p}K^-}.     \label{angular3}
\end{equation}
We shall be able to use this equality as a test of the penguin 
dominance. The corresponding relation holds between 
$B^0\to p\overline{p}K^0$ and $\overline{B}^0\to p\overline{p}
\overline{K}^0$. 

  At this stage we can prove that the $p\overline{p}$ enhancement 
is not a resonance, for instance, a glueball\cite{Glueball}: The Wigner 
functions $d^J_{0\lambda}(\theta)$, which are proportional to the 
associated Legendre functions, possess a special symmetry property 
under $\theta\leftrightarrow \pi-\theta$\cite{Rose}, 
\begin{equation}
 d^J_{0\lambda}(\pi-\theta) = (-1)^{J+\lambda}d^J_{0\lambda}(\theta)
                   \;\; \to \;\;
  |d^J_{0\lambda}(\pi-\theta)|^2 = |d^J_{0\lambda}(\theta)|^2.
                        \label{Wigner1}
\end{equation}
If the $p\overline{p}$ pair is produced entirely through a resonance, 
only the term of the resonance spin $J$ contributes in Eq. (\ref{angular2})
without sum over $J$. Since the function $|d^J_{0\lambda}(\theta)|^2$ 
is unchanged under $\theta \to \pi-\theta$ ({\em i.e.,} 
$\cos\theta\to -\cos\theta$), so is $d\Gamma/d\Omega_p$ in this case. 
However, experiment shows a pronounced asymmetry between two hemispheres 
of $\cos\theta >0$ and $\cos\theta <0$. (Fig. 3.) In terms of the 
forward-backward asymmetry parameter\cite{PL},
\begin{equation} 
 A\equiv(N_{+}-N_{-})/(N_{+}+N_{-})=0.59_{-0.07}^{+0.08}
\end{equation} 
contrary to $A=0$ in the case of a single $J$. Although interference 
between the resonant and nonresonant amplitudes of different $J$ can 
produce some asymmetry in principle, such interference should be 
insignificant under the normal circumstance where the resonant amplitude
acquires the phase $\frac{\pi}{2}$ through the resonant decay relative
to the nonresonant amplitude: $\arg(A^J_{{\rm res}}A^{J'*}_{{\rm non}})
\simeq\pm\pi/2$. It is implicitly assumed here as usual that the 
nonresonant production amplitude dose not acquire a significant phase. 
If for some reason the large asymmetry of Fig.3 should be caused by the
interference between the resonant and nonresonant amplitudes of 
different $J$'s, the very small yield observed toward $\theta_p=\pi$ in 
Fig. 3 would mean nearly perfect destructive interference between them. 
In this case the nonresonant yield would have to be just as large as the 
resonant one. Therefore the marked asymmetry in the angular correlation
rules out convincingly the hypothesis of pure resonant production. If one 
attempts to explain the the enhancement by strong FSI in a single 
dominant partial-wave channel of $p\overline{p}$\cite{Meissner},
one would likewise obtain $A\simeq 0$ for the angular correlation. 
To be consistent with the observed angular correlation, partial-wave 
amplitudes of even and odd $J$ must coexist and almost maximally 
interfere. Our argument is very general and independent of dynamics
up to this point. We now proceed to take dynamics into account. 

\noindent
\begin{figure}[h]
\hskip 3cm
\epsfig{file=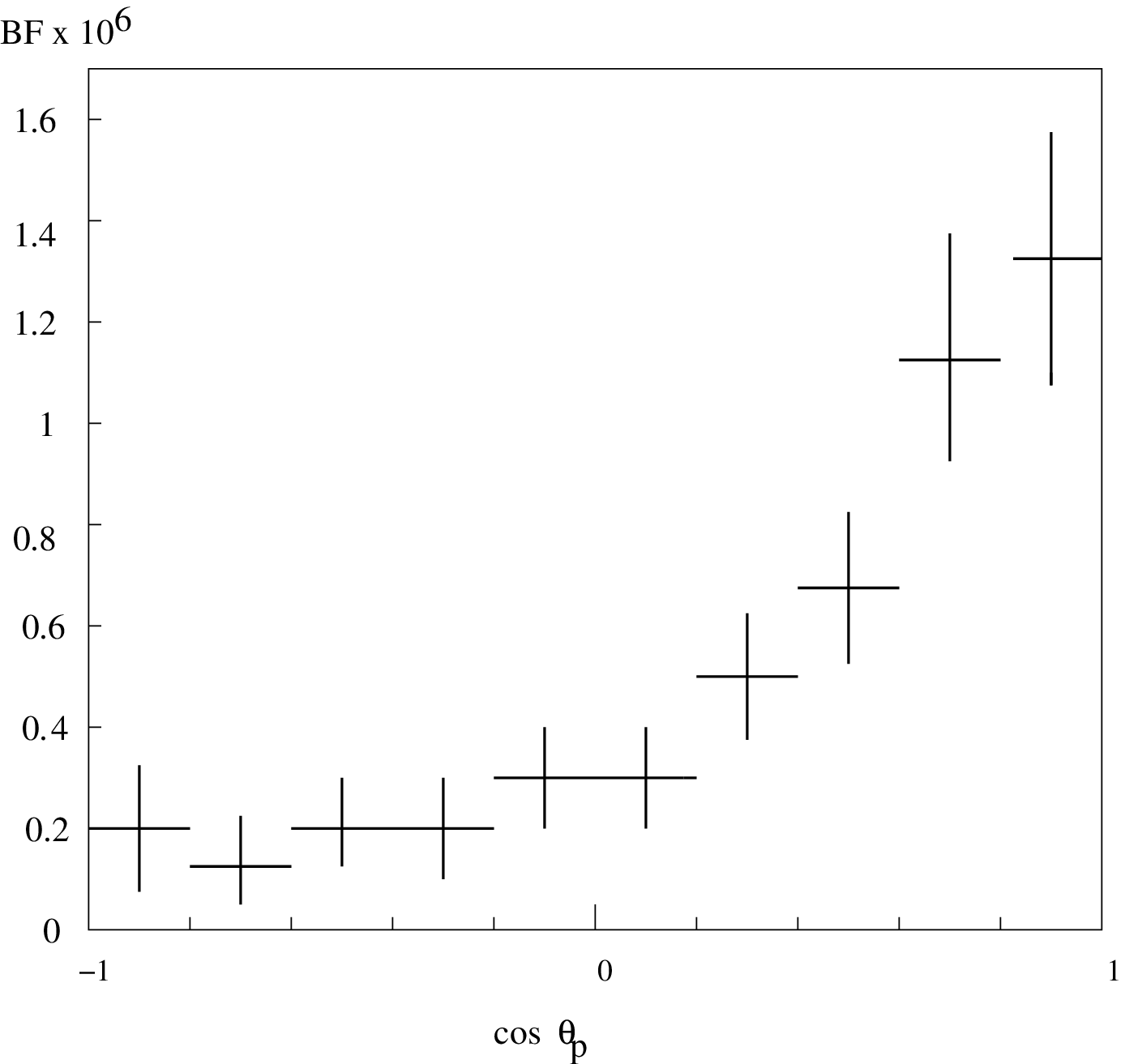,width=7cm,height=7cm}
\caption{The $pK^-$ angular distribution in the 
$p\overline{p}$ rest frame [Ref. 10].
\label{fig:3}} 
\end{figure}

In the SD picture the spectator $\overline{u}$-quark of $B^-$ enters 
the antiproton with two energetic antiquarks ($\overline{u}$ and 
$\overline{d}$) which are pair-produced nearly collinearly by two gluons. 
(Fig. 1b) Note that it is a color-suppressed process for the $s$-quark 
from $b\to sg^*$ of the strong penguin decay to form $K^-$ by capturing 
the spectator $\overline{u}$.\footnote{
We leave out the radiative penguin interaction here since it does not
affect the leading behavior due to the strong penguin interaction.} 
In contrast, the proton consists of three energetic quarks; 
one from the primary decay interaction and two of pair-produced quarks. 
In the $B^-$ rest frame, therefore, the proton recoils against $K^-$ more 
energetically on average than the antiproton does. It means that in the 
$p\overline{p}$ rest frame, the proton tends to move away from $K^-$ 
faster than the antiproton does. That is, $A < 0$ contrary to the 
measurement. This is the ``angular correlation puzzle''. No reasonable 
explanation has been given from the SD viewpoint. We must look for some 
LD interaction effect that has not been commonly appreciated.

   The maximum of the $p\overline{p}$ enhancement occurs near 
$m_{p\overline{p}}=$ 2 GeV in the BaBar data\cite{BaBar} and roughly
$\leq$2.2 GeV in the Belle data\cite{PL}. The dominant relative 
orbital angular momenta of $p\overline{p}$ are expected to be $s$-wave 
and $p$-wave. The amount of $d$-wave is presumably small and higher 
waves are even smaller. The terms that contribute dominantly 
in Eq. (\ref{angular1}) are therefore $J=0$ ($^1\!S_0, ^3\!P_0$), 
$J=1$ ($^3\!S_1, ^3\!P_1, ^1\!P_1$), and $J=2$ ($^3\!P_2$).
The explicit forms of the relevant $d^J_{0\lambda}$ functions ($J\leq 2$, 
$\lambda=-1,0,+1$) are\cite{Rose}: 
\begin{eqnarray}
  d^0_{00}(\theta) &=& 1, \;\;d^1_{00}(\theta) =\cos\theta,\;\;
  d^2_{00}(\theta) = (3\cos^2\theta -1)/2,\nonumber \\
  d^1_{0\pm 1}(\theta) &=& \mp\sqrt{1/2}\sin\theta,\;\;
  d^2_{0\pm 1}(\theta) = \mp\sqrt{3/2}\sin\theta\cos\theta,
                          \label{Wigner2}
\end{eqnarray}
It is convenient to rearrange the helicity decay amplitudes 
$A^{J}_{\lambda_p\lambda_{\overline{p}};0}$ with the spectroscopic 
notation into $A(^{2S+1}L_J;\lambda_p-\lambda_{\overline{p}})$. 
When only $s$-waves and $p$-waves 
are retained, the helicity amplitudes of definite isospin $I$ for
$N\overline{N}$ can be written as  
\begin{eqnarray}
   A^{0I}_{\pm\pm;0} &=&
    \pm A^I(^1S_0,0) + A^I(^3P_0,0), \nonumber \\
   A^{1I}_{\pm\pm;0} &=&
     A^I(^3S_1,0) \pm A^I(^1P_1,0), \nonumber \\
   A^{1I}_{\pm\mp;0} &=&
      \sqrt{2}A^I(^3S_1;\pm 1)\pm A^I(^3P_1;\pm 1),
          \nonumber \\
   A^{2I}_{\pm\pm;0} &=& A^I(^3P_2;0), \nonumber \\
   A^{2I}_{\pm\mp;0} &=& \sqrt{3/2} A^I(^3P_2;\pm 1), \label{breakup}
\end{eqnarray}
where we have denoted the helicity indices $\lambda_p, 
\lambda_{\overline{p}}=\pm 1/2$ simply by $\pm$. 
We shall use this notation hereafter. The normalization of the 
amplitudes is arbitrary for the decay amplitudes which have no
unitarity constraint. Since $\Delta I=0$ for the strong penguin 
decay, the decay amplitudes for the charge eigenstates of $\overline{B}$ 
and $\overline{K}$ are given by the decay amplitudes $A^I(^{2S+1}L_J,\lambda)$
of definite $N\overline{N}$ isospin as
\begin{eqnarray}
  A^{K^-}(^{2S+1}L_J,\lambda) &=& 
 \sqrt{1/2}\Bigl[A^1(^{2S+1}L_J,\lambda)-A^0(^{2S+1}L_J,\lambda)\Bigr], \nonumber \\
  A^{\overline{K}^0}(^{2S+1}L_J,\lambda) &=& 
 \sqrt{1/2}\Bigl[A^1(^{2S+1}L_J,\lambda)+A^0(^{2S+1}L_J,\lambda)\Bigr].
            \label{isospin}
\end{eqnarray} 
Combining Eqs. (\ref{breakup}) and (\ref{isospin}), we obtain the
decay amplitudes as functions of $\theta_p$, $\phi_p$ and 
$m_{p\overline{p}}$.  The decay amplitudes for $B^+/B^0\to p
\overline{p}K^+/p\overline{p}K^0$ are obtained from those of 
$B^-/\overline{B}^0\to p\overline{p}K^-/p\overline{p}\overline{K}^0$ 
with the interchange $\lambda_p\leftrightarrow\lambda_{\overline{p}}$ 
followed by $\theta_p\to\pi-\theta_p$ and $\phi_p\to\pi+\phi_p$
up to the overall CP phase factor of the penguin decay.

We are now able to write the complete differential decay rate for
$B^-\to p\overline{p}K^-$ with $p\overline{p}$ in $s$ and $p$-waves
in the notation of $^{2S+1}L_J$:
\begin{eqnarray}
  \frac{d\Gamma}{dm_{p\overline{p}}d\Omega_p}\biggr|_{B^-\to p\overline{p}K^-}
    &=& \Gamma_0 \Bigl|\Bigl(A^{K^-}(^1S_0,0)+A^{K^-}(^3P_0,0)\Bigr) 
    +\Bigl(A^{K^-}(^3S_1,0)+A^{K^-}(^1P_1,0)\Bigr)\cos\theta_p \nonumber\\
   &+&A^{K^-}(^3P_2,0)(3\cos^2\theta_p-1)/2 \Bigr|^2 \nonumber\\
 &+&\Bigl|\Bigl(-A^{K^-}(^1S_0,0)+A^{K^-}(^3P_0,0)\Bigr)+
     \Bigl(A^{K^-}(^3S_1,0)-A^{K^-}(^1P_1,0)\Bigr)\cos\theta_p \nonumber\\
   &+&A^{K^-}(^3P_2,0)(3\cos^2\theta_p-1)/2 \Bigr|^2 \nonumber \\
 &+&\Bigl|\Bigl(A^{K^-}(^3S_1,1)+\sqrt{1/2}A^{K^-}(^3P_1,1)\Bigr)\sin\theta_p
                         \nonumber \\ 
 &+& A^{K^-}(^3P_2,1)(3\sin\theta_p\cos\theta_p)/2\Bigr|^2
          \nonumber \\
 &+&\Bigl|\Bigl(A^{K^-}(^3S_1,-1)-\sqrt{1/2}A^{K^-}(^3P_1,-1)\Bigr)\sin\theta_p 
                         \nonumber \\
 &+& A^{K^-}(^3P_2,-1)(3\sin\theta_p\cos\theta_p)/2\Bigr|^2,     
  \label{rate} 
\end{eqnarray}
where the $\phi_p$ dependence goes away from the squared amplitudes of
definite helicity $\lambda$. Before going further, we point out that the 
$s$-wave amplitudes alone cannot generate the asymmetric angular 
correlation for $B^-\to p\overline{p}K^-$ even though two $s$-wave 
amplitudes ($^1S_0$ and $^3S_1$) enter the right-hand side of 
Eq. (\ref{rate}): The reason is that the interference terms cancel out 
between $^1S_0$ and $^3S_1$ as
\begin{eqnarray}
   \frac{d\Gamma}{dm_{p\overline{p}}d\Omega_p}\biggr|_{B^-\to p\overline{p}K^-} 
       &=& \Gamma_0
    \Bigl[|A^{K^-}(^1S_0,0) +A^{K^-}(^3S_1,0)\cos\theta_p|^2 \nonumber \\
        &+&|-A^{K^-}(^1S_0,0) +A^{K^-}(^3S_1,0)\cos\theta_p|^2 \nonumber\\  
        &+& |A^{K^-}(^3S_1,1)|^2\sin^2\theta_p
        + |A^{K^-}(^3S_1,-1)|^2\sin^2\theta_p\Bigr],
\end{eqnarray}
and consequently $d\Gamma/d\Omega_p$ turns out to be symmetric under 
$\cos\theta_p\to-\cos\theta_p$.
The same statement holds valid for $p$-waves alone. The observed steep asymmetry 
(Fig. 3) requires more than one orbital angular momentum, most likely 
$s$-wave and $p$-wave. It is very important experimentally to study how the 
angular correlation varies as $p$-waves increase with $m_{p\overline{p}}$ 
relative to $s$-waves across the threshold enhancement. It does not make sense 
to make a theoretical fit to the shape of the $m_{p\overline{p}}$ plot 
without large interference between different $p\overline{p}$ partial-waves.   

The experimental uncertainty in the angular measurement limits 
quantitative analysis at present. Let us be content with qualitative
analysis in this paper by approximating or interpreting for simplicity 
the angular correlation in  Fig. 3 as $\sim (1+\cos\theta_p)^2$. 
This $\cos\theta_p$ dependence is realized if
\begin{equation}
      A^{K^-}(^1S_0,0) \simeq  A^{K^-}(^3S_1,0) 
   \simeq A^{K^-}(^1P_1,0) \simeq A^{K^-}(^3P_0,0), \;\;{\rm(Exp)} 
 \label{Exp1}
\end{equation}
and all other amplitudes are negligible. An alternative solution is
\begin{equation}
      A^{K^-}(^1S_0,0) \simeq - A^{K^-}(^3S_1,0)
   \simeq A^{K^-}(^1P_1,0)\simeq - A^{K^-}(^3P_0,0),\;\;{\rm(Exp)}
\label{Exp2}
\end{equation}
and all others are negligible. A small amount of $A^{K^-}(^3P_2,0)$ with
the same sign as $A^{K^-}(^3S_1,0)$ would improve the fit a little
by lowering the curve near $\cos\theta_p=0$ and raising it near 
$\cos\theta_p=\pm 1$, but it is not crucial to the essence of our 
qualitative argument. While an accurate prediction is difficult 
because of our deficiency in knowledge of the quark distribution 
in baryons, the SD argument predicts, as we have argued above, the 
sign of slope opposite to Fig. 3: The angular dependence should be more
like $(1-|a|\cos\theta_p)^2$ ($|a|\leq 1$) in the SD argument. This angular 
dependence corresponds to the partial-wave amplitudes, 
\begin{equation}
  A^{K^-}(^1S_0,0) \approx  A^{K^-}(^3S_1,0)
  \approx -A^{K^-}(^1P_1,0)  \approx -A^{K^-}(^3P_0,0), \;\;{\rm(SD)} 
           \label{SD1}
\end{equation}
or alternatively,
\begin{equation}
      A^{K^-}(^1S_0,0)  \approx - A^{K^-}(^3S_1,0)
    \approx - A^{K^-}(^1P_1,0)  \approx A^{K^-}(^3P_0,0), \;\;{\rm(SD)} 
           \label{SD2}
\end{equation}
instead of Eqs. (\ref{Exp1}) or (\ref{Exp2}). Comparing the SD
prediction with experiment, we find that the relative signs of the 
$s$-to-$p$-wave amplitudes are opposite. There are several 
alternatives that can alter the SD prediction in line with experiment:
Sign reversal of the $^1S_0$ and $^3S_1$ amplitudes brings Eq. (\ref{SD1})
to Eq. (\ref{Exp1}) and Eq. (\ref{SD2}) to Eq. (\ref{Exp2}). Alternatively,
sign reversal of $^1S_0$ and $^3P_0$ brings Eq. (\ref{SD1}) 
to Eq. (\ref{Exp2}) and Eq. (\ref{SD2}) to Eq. (\ref{Exp1}). Sign
reversal of $^3S_1$ and $^1P_1$ also accomplishes the same.   
We ask what LD effect can possibly cause the sign reversal from 
Eq. (\ref{SD1}) or (\ref{SD2}) to Eq. (\ref{Exp1}) or (\ref{Exp2}).   
In the next section we argue that the desired sign reversal may occur if 
bound states exist in some of the $p\overline{p}$ channels. Unlike 
the argument that has ruled out a $p\overline{p}$ resonance, this is 
speculative and admittedly less clean part of our argument. 

\section{Final-state interaction}

 The three-body final-state interaction {FSI} was analysed in the 
approximation of sum of two-body FSI since going beyond is mathematically
formidable\cite{Faddeev}. Fortunately, in the particle configuration of 
our interest where the invariant mass of $p\overline{p}$ is small and 
the $K$ meson recoils fast against $p\overline{p}$, it is a good 
approximation and at least a common practice to separate the two-body 
FSI of $p\overline{p}$ ignoring the rest of FSI. Inclusion of 
$p\overline{p}$ annihilation channels is more a difficult problem. 
If one wants to make a quantitative analysis, this will be a main source 
of uncertainty.\footnote{
The FSI of the $p\overline{p}$ was recently studied for an enhancement 
in $J/\Psi\to\gamma p\overline{p}$\cite{nuclear,nuclear2,nuclear3}. 
In this process, $m_{p\overline{p}}$ is even closer to the threshold 
and consequently the coulombic FSI may be relevant.} 
Our task here is not to obtain numerically accurate results but to 
search a possible cause of sign flip for the amplitudes in the FSI. 
In order to make the sign flip argument plausible, we do not need much 
more than a basic argument of the elastic two-body FSI and its 
diagrammatic explanation.
 
The standard practice in FSI resorts to potential theory and 
incorporates FSI by modifying the decay amplitudes with  
final particle rescattering as\cite{Gillespie}
\begin{equation}
          A^{JI}(s) \to  A^{JI}(s)/f^{JI}(-k),  
\end{equation}     
where $f^{JI}(k)$ stands for the Jost function\cite{Jost} of a partial-wave
eigenchannel in variable $k=\frac{1}{2}\sqrt{s-4m_N^2}$
($s=(p_p+p_{\overline{p}})^2$). It is normalized to $f^{JI}(\infty)=1$. 
This FSI factor sums up ladders or bubbles of final particle rescattering 
in potential. The Jost function can be expressed with the phase of 
scattering amplitude $\delta^{JI}$ in the Omn{\'e}s 
representation\cite{Omnes};
\begin{eqnarray}
 \frac{1}{f^{JI}(-k)}&=& e^{\Delta^{JI}(\nu)},   \nonumber \\
     \Delta^{JI}(\nu)&=&\frac{1}{\pi}\int_{\nu_0}^{\infty}
       \frac{\delta^{JI}(\nu')}{\nu'-\nu-i\epsilon}d\nu', \nonumber \\
   &=& \frac{{\cal P}}{\pi}\int_{\nu_0}^{\infty}
   \frac{\delta^{JI}(\nu')}{\nu'-\nu-i\epsilon}d\nu'+i\delta^{JI}(\nu),  
                 \label{FSI}
\end{eqnarray}        
where $\nu=k^2$. The lower bound $\nu_0$ of the dispersion integral 
is extended to the negative region ($s< 4m_N^2$) when $p\overline{p}$ 
annihilation into meson channels is taken into account.

If annihilation and inelastic scattering are ignored, the phase 
$\delta^{JI}(\nu)$ would be equal to the phase shift of $N\overline{N}$ 
scattering according to the so-called Watson's theorem\cite{Watson}. 
If there is a resonance in this elastic case, the phase shift 
$\delta^{JI}(\nu)$ rises from zero, passes through 
$\pi/2$ at the resonance ($\nu=\nu_R$) and approaches $\pi$ as
$\nu\to \infty$. (Fig. 4.) Therefore the phase of the decay amplitude 
acquires a minus sign ($=e^{i\pi}$) above the resonance $\nu = \nu_R$. 

\noindent
\begin{figure}[h]
\hskip 3cm
\epsfig{file=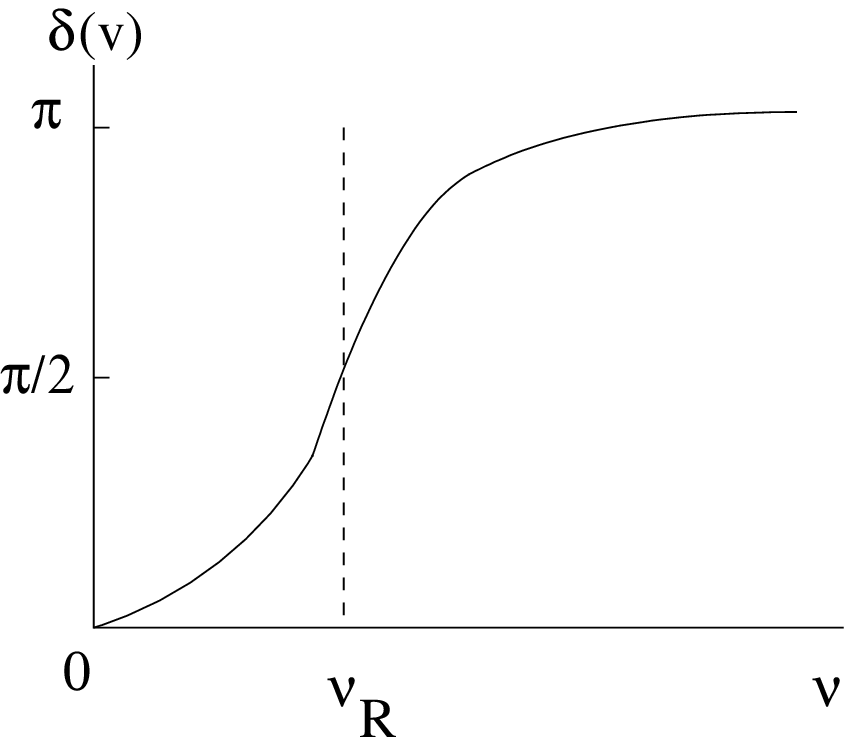,width=5cm,height=4cm}
\caption{The phase $\delta(\nu)$ of the FSI factor $e^{\Delta^{JI}}$
across an elastic resonance.
\label{fig:4}} 
\end{figure}

This negative sign is easily understood in diagram.      
When a final particle pair is produced through a resonance, 
as depicted in Fig. 5 for $B^-\to p\overline{p}K^-$, the decay 
amplitude near the resonance takes the form of 
\begin{equation}
  A^{JI}(s)\simeq\overline{A}^{JI}(s)
         \frac{g^2(s)}{m_R^2-im_R\Gamma_R(s)- s},     \label{Res}
\end{equation} 
where $\overline{A}^{JI}(s)$ is the amplitude in the absence of 
a resonance, $m_R$ and $\Gamma_R(m_R^2)$ are the resonance mass and 
width, and $g^2(s)$ is {\em positive} at $s=m_R^2$. The reason for 
positivity of $g^2(m_R^2)$ is as follows: By the phase theorem of 
FSI\cite{Watson}, it holds that $\arg[A^{JI}(s)/\overline{A}^{JI}(s)]
=\delta^{JI}(s)$. Therefore, the phase of 
$A^{JI}(s)/\overline{A}^{JI}(s)$ must be equal to $+\pi/2$, not
$-\pi/2$, at the resonance peak. The phase of the resonant FSI factor 
in Eq. (\ref{Res}) must agree with this value at the resonance. The 
phase of $\arg[ig^2/m_R\Gamma(m_R^2)]$ is $+\pi/2$ at $s = m_R^2$ 
in agreement with the phase theorem if $g^2(m_R^2) > 0$. The function 
$g^2(s)$ is expected to be only mildly energy dependent even near
the threshold since the centrifugal factor $k^{l}$ of orbital 
motion resides in $A^{JI}(s)$ and $\overline{A}^{JI}(s)$, not in 
$g^2(s)$. Since $g^2(s)$ $(\simeq g^2(m_R^2))$ is positive, the sign of 
$A^{JI}(s)$ is the same as that of $\overline{A}^{JI}(s)$ below the 
resonance ($m_R^2-s \gg m_R\Gamma_R$), but turns opposite above the 
resonance ($s-m_R^2 \gg m_R\Gamma_R$). This is another way of seeing 
the sign and energy dependence of the FSI factor across a resonance.

\noindent
\begin{figure}[h]
\hskip 3cm
\epsfig{file=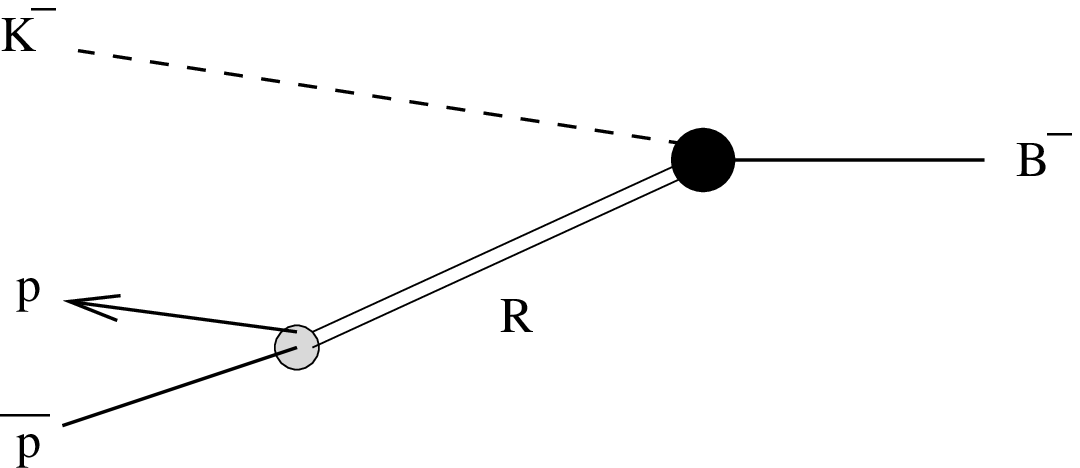,width=4.5cm,height=2cm}
\caption{Resonant production of $p\overline{p}$
in $B^-\to p\overline{p}K^-$.
\label{fig:5}} 
\end{figure}
 
This simple argument is modified by inelasticity above the 
$N\overline{N}\pi$ threshold and by annihilation into meson channels. 
Above the energies where inelastic channels start contributing 
substantially at $\sqrt{s}>2m_N+m_{\pi}$, the FSI formulas of potential 
theory is no longer applicable. If we simply truncate the phase integral 
in Eq. (\ref{FSI}) at $\nu=\nu_{\rm max}$ somewhere above the inelastic 
threshold, the FSI factor computed in 
the narrow-width (step-function) approximation turns out to be 
\begin{equation}
    e^{\Delta^{JI}(\nu)}\simeq \frac{\nu_{\rm max}-\nu}{\nu_R -\nu},
                    \;\;\; (\nu \gg \nu_R),
\end{equation}
which satisfies $f(-k)\to 1$ as $\nu(=k^2)\to \infty$. This FSI 
factor is negative between the resonance and the inelastic threshold;
\begin{equation}
  1/f(-k)= e^{\Delta^{JI}(\nu)} < 0, \;\;\; 
                (\nu_R < \nu < \nu_{\rm max}). \label{sign} 
\end{equation}
It means that the FSI factor gives a minus sign above the resonance 
until energy goes up so high that inelasticity becomes important. 
While the negative sign is easy to understand, magnitude of the FSI 
factor is harder to estimate since it depends on the dispersion integral 
over the entire energy range. 

Let us turn to the effect of the annihilation channels into mesons. 
The first issue is that the phase $\delta^{JI}$ of the decay amplitude 
is no longer equal to the phase of the $N\overline{N}$ scattering 
amplitude at any energy where annihilation occurs. An approximate equality 
between two phases holds only in those eigenchannels in which the channel 
coupling is weak between $N\overline{N}$ and the meson channels. It 
is not obvious whether this is the case for the relevant $N\overline{N}$ 
channels near the threshold. We must assume it here. Discussion will 
be made on this point below and in the next section. The other 
issue is whether relevant $N\overline{N}$ resonances really exist 
or not. The candidates of $N\overline{N}$ bound states and resonances 
indeed exist.  Since the $N\overline{N}$ bound states can be only loosely 
bound, $N$ and $\overline{N}$ are spatially separated outside the range 
of annihilation interaction and therefore the annihilation into mesons is 
suppressed. Meanwhile, being a bound state, the state cannot decay into
a nucleon and an antinucleon since its mass $m_B$ is below the 
$N\overline{N}$ threshold ($m_B < 2m_N$). However, there is an escape 
from this argument: The finite lifetime due to meson annihilation 
generates a width to the mass of the bound-state by the time-energy 
uncertainty. If this width is a little wider than the binding energy 
$\Delta = 2m_N-m_B$, the $N\overline{N}$ ``bound state'' can decay into 
$N\overline{N}$ (Fig. 6). This decay suffers a severe phase space 
suppression. When the $N\overline{N}$ bound state is produced 
``on mass shell'' ($2m_N<M_{N\overline{N}}< 2m_N+\Gamma$ with 
$\Gamma$ being the width), its decay branching fraction to the 
$N\overline{N}$ channel is small even if its coupling to $N\overline{N}$ 
is strong, {\em i.e.,} even if $g^2(s)$ in Eq. (\ref{Res}) is large.  
Consequently, such an $N\overline{N}$ bound state would appear as 
a relatively narrow meson resonance. On the other hand, when an 
$N\overline{N}$ pair is produced above the width of the bound state, 
the bound state can still enhance $N\overline{N}$ production through the 
small denominator of the resonance propagator. In experiment the transition 
from ``on-shell'' to ``off-shell'' occurs continuously above the threshold. 
The phase-space factor pushes the enhancement peak upwards from $2m_N$ to 
$m_{N\overline{N}} = 2m_N+O(\sqrt{\Delta^2+\Gamma^2/4})$. When 
the phase-space factor is removed, the yield curve is expected to 
behave like $g^2(s)/[(m_{N\overline{N}}-m_B)^2+\Gamma^2/4]$ at 
$m_{N\overline{N}}> 2m_N$. The BES Collaboration\cite{BES} first extracted 
the resonance parameters on this assumption when analysis was made only 
above the $p\overline{p}$ threshold in $J/\psi\to p\overline{p}\gamma$. 
In the three-body baryonic $B$ decay, the events of small 
$m_{N\overline{N}}$ receive the SD enhancement, as we have argued.
The same SD effect would be less prominent in $J/\psi$ decay since the 
phase space is much smaller. Consequently the location and the shape of 
the enhancement may not be identical in $B$ and $J/\psi$ decays. 
Magnitude of the net $p\overline{p}$ enhancement is also dependent on  
dynamical environment of production. 
Despite such dynamical uncertainties we are fairly confident that 
if enhancement indeed occurs in the region of $m_{N\overline{N}}>2m_N$, 
and if it couples to a state below it, the decay amplitude acquires the 
negative sign of FSI according to the diagram in Fig. 5 and the discussion 
following Eq. (\ref{Res}).  
\noindent
\begin{figure}[h]
\hskip 3cm
\epsfig{file=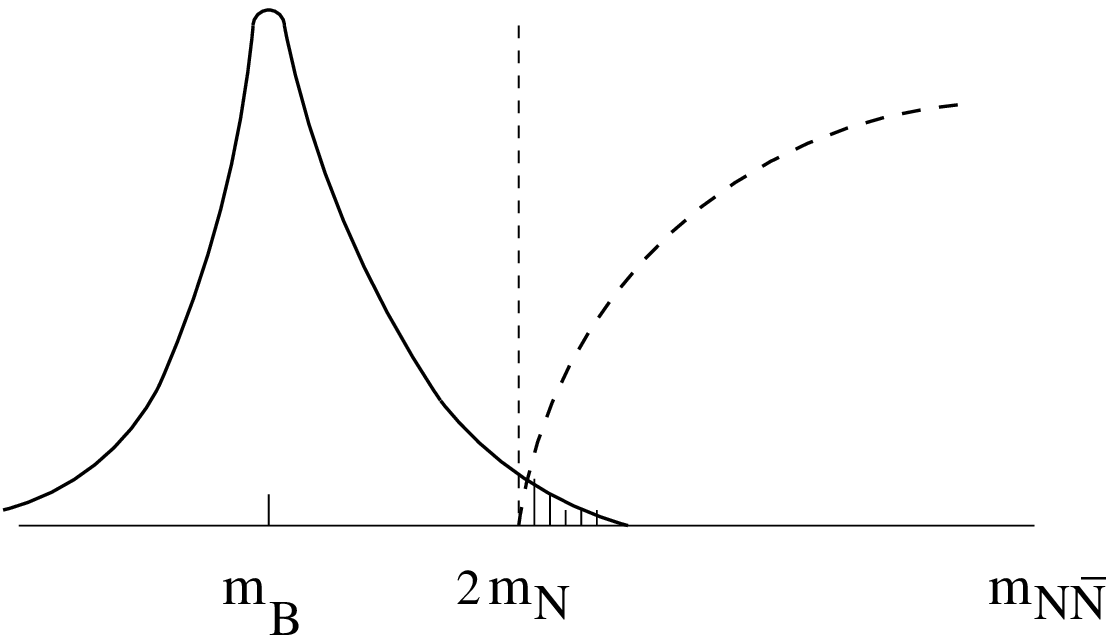,width=5cm,height=3.5cm}
\caption{The $N\overline{N}$ bound state $X$ with mass $m_B$ acquires 
a small width by annihilation decay. When the $X$ is produced
 ``on mass shell'', only the upper corner (the hatched region) above the 
$N\overline{N}$ threshold contributes to the decay $X\to N\overline{N}$.
The broken curve above $2m_N$ indicates the $s$-wave phase space.
The $N\overline{N}$ production above the $X$ mass shell can still be 
significant by the enhancement due to the $X$ propagator. 
\label{fig:6}} 
\end{figure} 
   
We remark on the coupling between $N\overline{N}$ and the annihilation 
channels. In the $p\overline{p}$ reaction at the threshold the annihilation 
cross section is larger than the elastic scattering cross section. Can it 
be compatible with weak coupling between $N\overline{N}$ and annihilation 
channels ?  We should first note that the annihilation cross sections 
fall very rapidly with the inverse flux factor 
$1/|{\bf v}_p-{\bf v}_{\overline{p}}|$ above the threshold according 
to the ``$1/v$'' law of the exothermic reactions.
We should also note that the large annihilation cross section
is largely due to multitude of multi-meson annihilation channels 
with many different partial waves of subchannels.  In contrast, the 
elastic cross section near the threshold is almost entirely due to 
$s$-wave scattering.  The annihilation cross section may not be so 
large in many partial-wave eigenchannels a little above the threshold.  
Therefore, the experimentally observed large total annihilation 
cross section is not an outright contradiction with weak coupling 
between $N\overline{N}$ and meson channels.

Theorists are not unanimous about existence of the $N\overline{N}$ 
bound states and resonances\cite{nuclear,nuclear2,nuclear3}. 
The recent discovery of the state $X(1835)$ in the 
radiative $J/\psi$ decay suggests that an $N\overline{N}$ bound state 
may exist after all. If $X(1835)$ is indeed an $N\overline{N}$ bound 
state in $^1S_0$ or $^3P_0$, it is conceivable that a $N\overline{N}$ 
bound state exists in the $^3S_1$, $^3P_1$ or $^1P_1$ channel as well. 
Because of negative charge parity, experimental search is harder for 
$^3S_1$ and $^1P_1$ in the radiative $J/\psi$ decay than search of 
$^1S_0$ and $^3P_J$. Leaving existence of $N\overline{N}$ bound states 
as an experimental issue still open, we proceed with our hypothesis 
of the sign flip and study the consequences in the $pK^-$ angular 
correlation.

\section{$N\overline{N}$ bound states and $p\overline{K}$ angular correlation}

The $\pi\pi\eta'$ resonance $X(1835)$ is the best candidate for the 
$N\overline{N}$ bound state. The sharp $p\overline{p}$ threshold enhancement
observed in $J\psi\to p\overline{p}\gamma$ first hinted its existence as an
$N\overline{N}$ bound state\cite{BES}. The mass was deduced at $1859_{-10}^{+3}
({\rm stat})_{-25}^{+5}({\rm sys})$ MeV with width $< 30$ MeV. These values
are sensitive to the method and assumptions involved in extracting them, 
{\em e.g.} rescattering and nonresonant background. They called it $X(1859)$. 
Two years later the BES Collaboration\cite{Ablikim} identified a resonance 
in the $\pi\pi\eta$ mass plot and called it $X(1835)$, which is presumably 
the same state as $X(1859)$. It is most likely a state of $^1S_0$ with 
$I=0$\cite{Ablikim}. Assignment to $^3P_0$ ($\sigma$ and $\eta'$ in $p$-wave) 
of $I=0$ cannot be excluded purely experimentally though less likely in 
theory because of the centrifugal repulsion. The width $(67.7\pm 20.3\pm 7.7)$ 
MeV is fairly narrow for its high mass. The upper tail of 
the width extends beyond the $p\overline{p}$ threshold and contributes to 
the decay into $p\overline{p}$. The BES Collaboration quotes the ratio of 
branching fractions as ${\rm Br}(X(1835)\to p\overline{p})/{\rm Br}
(X(1835)\to\pi^+\pi^-\eta')\simeq 1/3$. In view of the tiny $p\overline{p}$ 
phase space, we reason that coupling of $X(1835)$ to $p\overline{p}$ is 
much stronger than that to mesons.  For this reason the BES Collaboration 
suggests that $X(1835)$ is a likely candidate for a molecular or 
deuteron-like $N\overline{N}$ bound state. Such a bound state can play 
a dominant role in producing a $p\overline{p}$ pair in its eigenchannel 
near the threshold with little annihilation into mesons. This is exactly 
the state that we want for the cause of the sign flip. 

If an $N\overline{N}$ bound state exists in $^1S_0$, a bound state may 
exist in  $^3S_1$ as well by the property of the meson-exchange force 
between $N$ and $\overline{N}$. If so, the decay amplitudes $A^{K^-}(^1S_0,0)$ 
and $A^{K^-}(^3S_1,0)$ flip their signs from the SD ones in Eq. (\ref{SD1}) 
or Eq. (\ref{SD2}) to the experimentally observed ones in Eq. (\ref{Exp1})
or Eq. (\ref{Exp2}). In this way we would have a chance to obtain the 
observed trend $(1+\cos\theta_p)^2$ for the $pK^-$ angular correlation.
Since the $B^-\to p\overline{p}K^-$ amplitudes consist of both $I=0$ and 
$I=1$ of $N\overline{N}$, the sign flip would occur most effectively when the 
$I=0$ amplitudes dominate over the $I=1$ 
amplitudes;
\begin{equation}
     |A^0(^1S_0,0)|\gg |A^1(^1S_0,0)|,\;\;\;
     |A^0(^3S_1,0)|\gg |A^1(^3S_1,0)|.    \label{I=0}
\end{equation}
If the $^3S_1$ bound state is in $I=1$ instead of $I=0$, the second 
inequality in Eq. (\ref{I=0}) should be reversed in direction.

This is our proposal for the resolution of the angular correlation puzzle.
As was mentioned below Eq. (\ref{SD2}), there are other possibilities if
a $p$-wave bound state exists. In those cases the $s$-wave $N\overline{N}$
bound state should exist only in $^1S_0$ or $^3S_1$, not in both. The spin
splitting is generally weaker in nuclear forces than the orbital-angular-momentum
splitting. If this prevails in the $N\overline{N}$ force,
the bound states should appear first in the $s$-wave channels and then in
the $p$-wave channels. However, we should keep our mind open to the other
possibilities of sign reversal in the $p$-wave amplitudes.   
In order to make further advance, we need to know more about $X(1835)$ and
to search for more candidates of the $N\overline{N}$ bound states. In 
experiment of $B$ meson physics, we are anxious to know the $m_{p\overline{p}}$ 
dependence of the angular correlation since it will provide important pieces of
information about spin-parity, mass and isospin of the bound states. 

As for $B^0/\overline{B}^0\to p\overline{p}K_S$, the measurement was made 
for the oscillating $B^0$-$\overline{B}^0$ averaged over time and therefore
no flavor information is available\cite{PL}. So long as 
the penguin interaction dominates, the time-averaged $pK_S$ angular 
distribution is symmetric under $\cos\theta_p\to -\cos\theta_p$ in general.  
(cf Eq. (\ref{angular3}).) Specifically, if we keep only the amplitudes 
of $J\leq 1$ and $\lambda=0$ in Eq. (\ref{rate}), the angular distribution is 
\begin{eqnarray}
 \frac{d\Gamma}{dm_{p\overline{p}}d\Omega_p}\biggr|_{B^0/
      \overline{B}^0\to p\overline{p}K_S}&=&\Gamma_0\Bigl[
 \Bigl(|A^{\overline{K}^0}(^1S_0,0)|^2+|A^{\overline{K}^0}(^3P_0,0)|^2\Bigr)
           \nonumber \\
 &+& \Bigl(|A^{\overline{K}^0}(^3S_1,0)|^2+|A^{\overline{K}^0}(^1P_1,0)|^2
     \Bigr)\cos^2\theta_p\Bigr]. 
\end{eqnarray}
The curve of this angular correlation for $pK_S$ is concave in $\cos\theta_p$. 
The data\cite{PL} in Fig. 7 indeed show the tendency of roughly 
$\sim 1+\cos^2\theta_p$ albeit with very large uncertainty. The branching 
fraction was also measured\cite{PL} and its ratio to that of 
$p\overline{p}K^-$ is
\begin{equation}
{\rm Br}(B^0/\overline{B}^0\to p\overline{p}K_S)/
{\rm Br}(B^+\to p\overline{p}K^+) \simeq 0.23 \label{BrExpt} 
\end{equation}
with roughly $\pm 20$\% of statistical errors and $\pm 10$\% of systematic 
errors. This number would be 0.5 if the $I=0$ amplitudes completely dominates
over the $I=1$ amplitudes in Eq. (\ref{I=0}). If the I=1 amplitudes are
about 20\% of the $I=0$ amplitudes, however, this ratio 0.23 can be 
reproduced.

\noindent
\begin{figure}[h]
\hskip 3cm
\epsfig{file=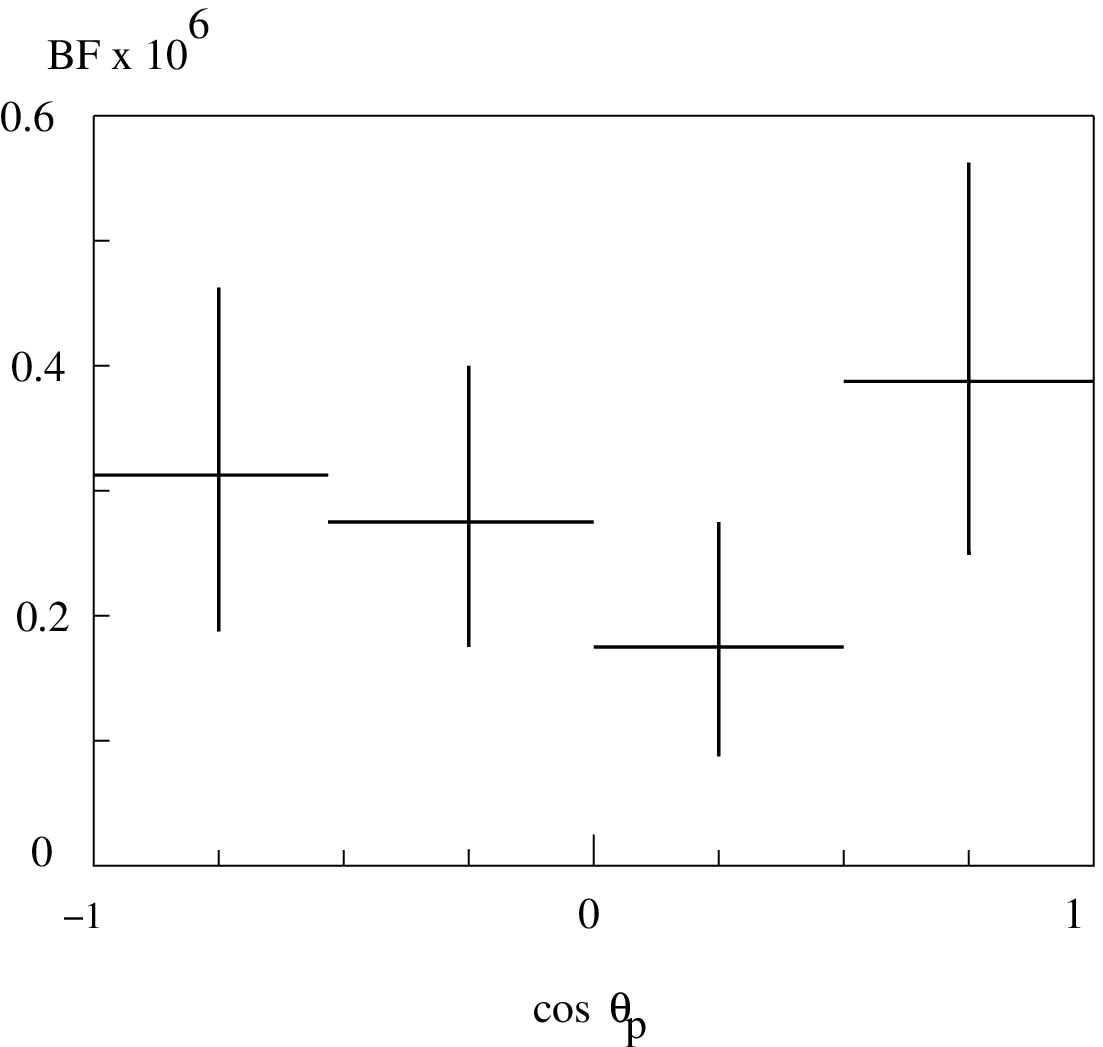,width=7cm,height=7cm}
\caption{The $pK_S$ angular correlation
in the $p\overline{p}$ rest frame of 
$B^0/\overline{B}^0\to p\overline{p}K_S$ [Ref. 10].
\label{fig:7}} 
\end{figure}

\section{$\Lambda\overline{p}$ channel}
   
The threshold enhancement has been observed in other three-body 
baryonic final states, $\Lambda\overline{p}\pi^+$,
$p\overline{p}\pi^-$, and $\Lambda\overline{\Lambda}K^+$ (charge
conjugated states combined) as well as in $\Lambda\overline{p}\gamma$ 
and many decay modes of the $b\to c$ transition. However, the angular 
correlation has been measured only for a few of them; 
$\Lambda\overline{p}\pi^+$, ${\Lambda}\overline{p}\gamma$, and 
${\Lambda}_c\overline{p}\pi^+$ all in low statistics. Let us look into
the $\Lambda\overline{p}$ enhancement observed in $\Lambda\overline{p}
\pi^+$ and $\Lambda\overline{p}\gamma$ of $\overline{B}^0$ decay and
the conjugate, for which the dominant process is the penguin decay.

 In the color-dominant process of $\overline{B}^0\to
\Lambda\overline{p}\pi^+$ the spectator $\overline{d}$-quark forms 
$\pi^+$ through capture by the energetic $u$-quark that comes directly from 
the strong penguin interaction. Since neither $\Lambda$ nor $\overline{p}$ 
picks up the slow spectator, their average energies in the 
$\overline{B}^0$ rest frame should be comparable. That is, the 
$\overline{p}\pi^+$ angular correlation in the $\Lambda\overline{p}$ 
rest frame ought to be more or less symmetric and flat in 
$\cos\theta_{\overline{p}}$. This naive SD prediction is in line 
with experiment within large uncertainty: The measured angular 
correlation\cite{PL} does not show marked asymmetry nor large 
variation in $\cos\theta_{\overline{p}}$. (See Fig. 8.) The 
$\overline{p}\pi^+$ angular distribution can be fitted with any 
of $1$, $1+|b|^2\sin^2\theta_{\overline{p}}$, and 
$(1-|b|\cos\theta_{\overline{p}})^2$ with small constant $|b|$. Although
the angular distribution is consistent with the SD prediction, one 
cannot rule out a resonance for the $\Lambda\overline{p}$ enhancement:  
A resonance of $J=0$ leading to $d\Gamma/d\Omega_{\overline{p}}\sim 1$ 
is certainly acceptable. A resonance of 
$J=1$ is neither ruled out since the flat distribution arises with  
$|A^{1}_{++;0}|^2+|A^{1}_{--;0}|^2\approx\frac{1}{2}(|A^{1}_{+-;0}|^2
+|A^{1}_{-+;0}|^2)$.

\noindent
\begin{figure}[h]
\hskip 3cm
\epsfig{file=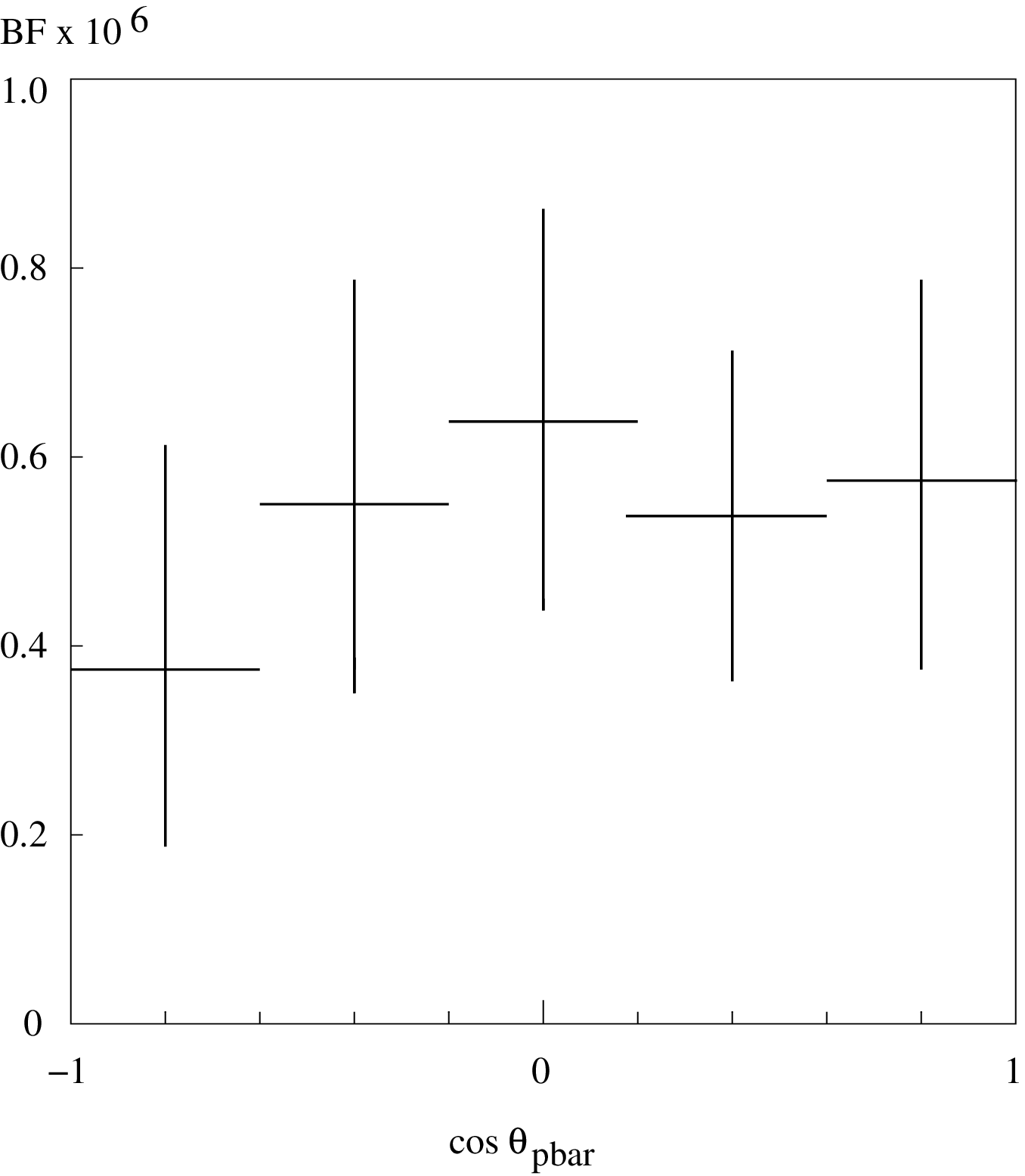,width=7cm,height=7cm}
\caption{The $\overline{p}\pi^+$ angular correlation 
in the $\Lambda\overline{p}$ rest frame of 
$\overline{B}^0\to\Lambda\overline{p}\pi^+$ [Ref. 10]. 
\label{fig:8}} 
\end{figure}
 
Let us turn to $\Lambda\overline{p}\gamma$. In the SD picture the 
energetic $s$-quark is emitted against $\gamma$ by the radiative 
penguin interaction and becomes the constituent of $\Lambda$. Other
quarks and antiquarks are produced through strong interaction nearly 
collinearly against $\gamma$ to avoid creating a fat gluon. Since
the spectator $\overline{d}$ enters in $\overline{p}$, the antiproton 
is less energetic in the rest frame of $\overline{B}^0$ than $\Lambda$.  
The SD prediction is therefore that $\overline{p}$ moves along
$\gamma$ in the $\Lambda\overline{p}$ rest frame, that is,  
$d\Gamma/d\Omega_{\overline{p}}$ should rise toward 
$\cos\theta_{\overline{p}}=1$, where $\theta_{\overline{p}}$ is 
the angle between $\overline{p}$ and $\gamma$ in the 
$\Lambda\overline{p}$ rest frame. The observed $\overline{p}\gamma$ 
angular correlation clearly shows this trend in line with the 
SD prediction. (See Fig. 9.)

\noindent
\begin{figure}[h]
\hskip 3cm
\epsfig{file=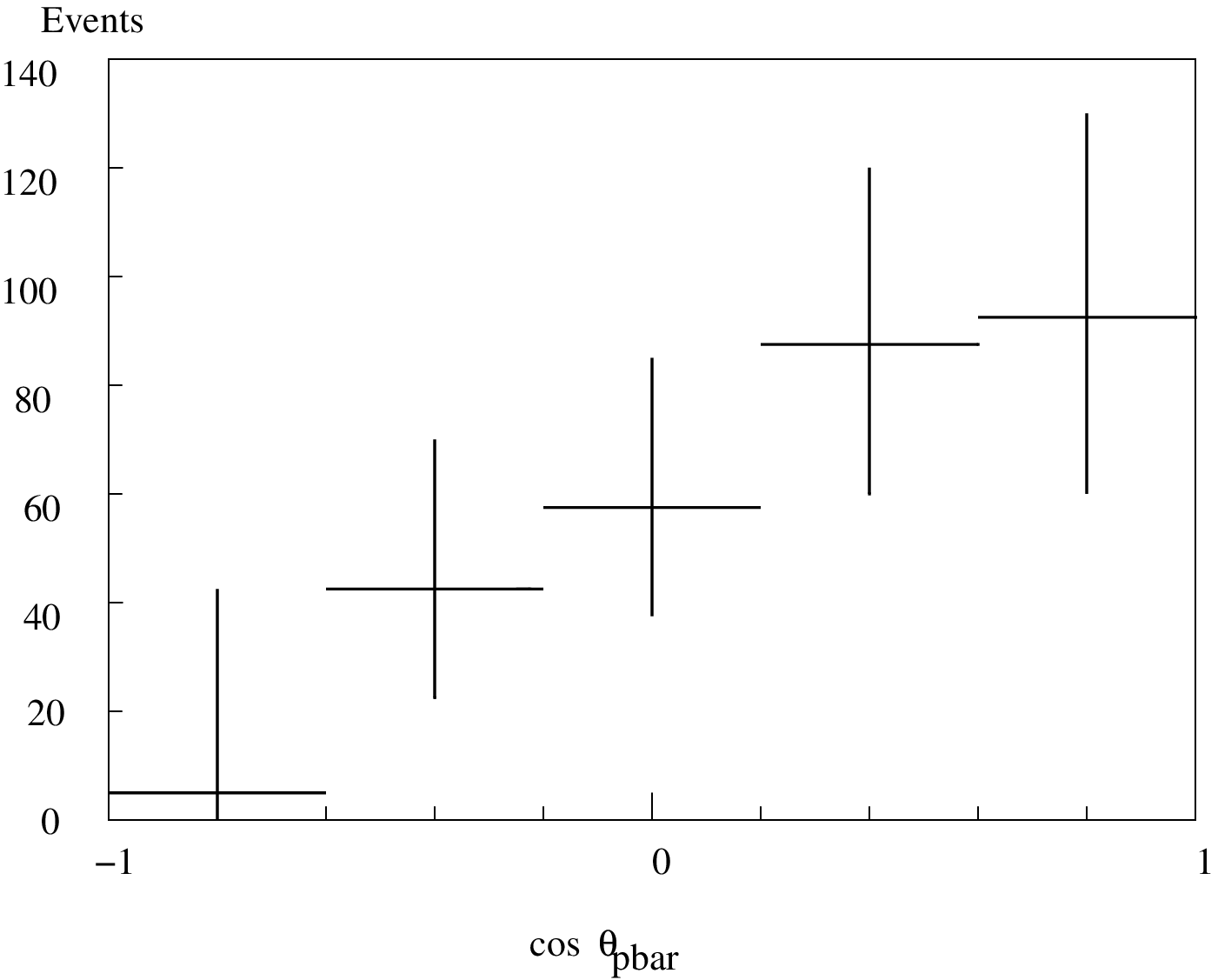,width=7cm,height=7cm}
\caption{The $\overline{p}\gamma$ angular correlation 
in the $\Lambda\overline{p}$ rest frame of 
$B^-\to \Lambda\overline{p}\gamma$ [Ref. 26]. 
\label{fig:9}} 
\end{figure}

Although the SD prediction is right for $\Lambda\gamma$, a resonance 
cannot be ruled out for $\Lambda\overline{p}$.  The helicity expansion 
of the $\Lambda\overline{p}\gamma$ amplitude is modified by spin of 
$\gamma$. The helicity of $\gamma$ is $-1$ since the $s$-quark emitted 
by the penguin interaction is left-handed. Therefore the component of
total angular momentum along the photon momentum 
${\bf J}\cdot{\bf{\hat{p}}}_{\gamma}$ is $+1$ for $\Lambda\overline{p}$ 
in their rest frame.  The $\Lambda\overline{p}$ angular correlation is 
given generally by
\begin{equation}
     \frac{d\Gamma}{dm_{\Lambda\overline{p}}d\Omega_p}
     \Biggr|_{B^-\to\Lambda\overline{p}\gamma} 
     =\Gamma_0\sum_{\lambda_{\Lambda}\lambda_{\overline{p}}}
 |\sum_J A^{J}_{\lambda_{\overline{p}}\lambda_{\Lambda};1}d^J_{1\lambda}
  (\theta_{\overline{p}})|^2, \;\;\; 
    (\lambda=\lambda_{\overline{p}}-\lambda_{\Lambda}).
\end{equation}
Note here that the first subscript of $d^J_{1\lambda}
(\theta_{\overline{p}})$ is $1$ owing to $J_z=+1$. The $\Lambda\overline{p}$ 
enhancement cannot be a resonance of $J=0$ since $J\geq |\mu|$ for 
$d^J_{\mu\lambda}(\theta_{\overline{p}})$. However, we cannot rule out 
a $J=1$ resonance since with  
\begin{equation}
  1+\cos\theta_{\overline{p}}
      = 2 d^1_{11}(\theta_p)^2 + d^1_{10}(\theta_p)^2, 
\end{equation}
the linear correlation $\sim 1+\cos\theta_{\overline{p}}$ arises
if the helicity amplitudes for $\Lambda\overline{p}\gamma$ happen to 
obey, for instance,
\begin{equation}
   |A^{1}_{+-;1}|^2\approx
  2\Bigl(|A^{1}_{++;1}|^2+ |A^{1}_{--;1}|^2\Bigr),\;\;
     A^{1}_{-+;1}\approx 0.       
\end{equation} 
Because of the preferred photon helicity, the $\overline{p}\gamma$
angular correlation can be asymmetric under $\cos\theta_{\overline{p}}
\to -\cos\theta_{\overline{p}}$ even with a single partial wave unlike 
those of $pK^-$ and $\overline{p}\pi^+$.  Combining our observations 
in $B^-\to\Lambda \overline{p}\pi^+$ and $B^-\to\Lambda\overline{p}\gamma$ 
together, we can rule out a $J=0$ resonance for $\Lambda\overline{p}$, 
but not a $J=1$ resonance. However, there is no motivation 
to call for a $\Lambda\overline{p}$ resonance or bound state at
present until we see clear discrepancy with the SD prediction. 
    
\section{Summary and remarks}
 
Three-body baryonic decay modes are favoured over two-body baryonic 
decay modes since a baryon-antibaryon pair may be emitted nearly 
collinearly in three-body decay. Indeed experiment confirms that 
the invariant mass of the baryon-antibaryon pair is strongly enhanced 
near the threshold in most modes. Although the SD picture appears to 
describe general trends of three-body decays in most cases, we have 
encountered one clear contradiction with the SD picture in the angular 
correlation between $p$ and $K^{\pm}$ in $B^{\pm}\to p\overline{p}K^{\pm}$. 

  Failure of the SD picture means that some LD effect enters the 
process of $B^-\to p\overline{p}K^-$ and reverses the angular 
dependence. We have pointed our finger to the nucleon-antinucleon 
bound states for the cause of sign flip of the SD amplitudes and have 
given a simple diagrammatic explanation for it. To explain the decay
angular correlation for $B^+\to p\overline{p}K^+$, we have postulated 
that $X(1835)$ be the 
$^1S_0$ bound state of $N\overline{N}$. That is, $X(1835)$ is a molecular 
six-quark state $qqq\overline{q}\overline{q}\overline{q}$ or a deuteron-like 
state and primarily couples to $N\overline{N}$ rather than to mesons.  
Many theorists have made the same or similar proposals on the nature 
of $X(1835)$, with motivations very different from ours. In addition, 
we need a $^3S_1$ bound state of $N\overline{N}$. The maximum asymmetry 
of the angular correlation should occur at the energy where kinematically
rising $p$-wave amplitudes become comparable with the falling $s$-wave 
amplitudes. The $m_{p\overline{p}}$ angular dependence will tell us 
a lot about dynamics near the $p\overline{p}$ threshold.

Our argument depends on strong interaction dynamics near the 
$N\overline{N}$ threshold that has not been proven nor disproven
experimentally. Some might feel that we have blown out a possible 
solution to a small puzzle into a farfetched speculation. We 
cannot counter such objections effectively. Our argument presented in 
this paper is a conjecture or a hypothesis, certainly not a theorem.  
Although our argument is only exploratory and speculative, the sign 
flip by a bound state or a resonance can occur generally and cause 
failure of the simple quark-gluon argument of multi-body $B$ meson decay.

While our argument of the sign flip is exploratory in nature, we would 
like to emphasize that the partial-wave expansion analysed here will be 
very useful as a general tool to penetrate into complexity of three-body 
decay dynamics. To show its usefulness, we have ruled out convincingly 
the resonance hypothesis for the $p\overline{p}$ threshold enhancement. 
We have also shown that the FSI in $s$-waves or in $p$-waves alone should 
not describe the enhancement either. While the angular correlation 
measurement is not extensive nor accurate enough at present, we expect 
that partial-wave analysis of three-body decay will shed more of new 
light on dynamics of $B$ decay in near future since the data are rapidly 
accumulating.

\acknowledgments

This work was supported by the Director, Office of Science, Office of
High Energy and Nuclear Physics, Division of High Energy Physics,
of the U.S.  Department of Energy under contract DE--AC02--05CH11231. 


\begin{thebibliography}{}

\bibitem{Belle1} K. Abe {\em et al.}, Belle Collaboration, 
{\em Phys. Rev. Lett.} {\bf 88}, 181803 (2003).
\bibitem{Belle2} M.Z. Wang {\em et al.,} Belle Collaboration, 
     {\em Phys. Rev. Lett.} {\bf 90}, 201802 (2003); {\bf 92}, 131801 (2004); 
     Y.J. Lee, Belle Collaboration,
     {\em Phys. Rev. Lett.} {\bf 93}, 211801 (2004).
\bibitem{BaBar} B. Aubert {\em et al.,} {\em Phys. Rev.} {\bf D 72} 051101 (2005).
\bibitem{Belle3} N. Gabyshev {\em et al.,} Belle Collaboration, {\em Phys. Rev. 
        D} {\bf 66}, 091102 (2002); K Abe {\em et al.,} Belle Collaboration,
        {\em Phys. Rev. Lett.} {\bf 89}, 151802 (2002); B. Aubert {\em et al.,}
       BaBar Collaboration, ICHEP 2004, Beijing, China, hep-ex/0408035; 
       N. Gabyshev {\em et al.,} Belle Collaboration, hep-ex/0409005.
\bibitem{Lee} Y. L. Lee {\em et al.,}, Belle Collaboration, {\em Phys. Rev. Lett.} 
       {\bf 95}, 061802 (2005).   
\bibitem{Soni} W-S. Hou and A. Soni, Phys. Rev. Lett. {\bf 86}, 4247 (2001).
\bibitem{Glueball} C. K. Chua, W. S. Hou, and S. Y. Tsai, {\em Phys. Lett.} 
       {\bf B 544}, 139 (2002).
\bibitem{TH} J. L. Rosner, {\em Phys. Rev.} {\bf D 68}, 014004 (2003). 
\bibitem{pole}  H. Y. Cheng and K. C. Yang, {\em Phys. Rev.} {\bf D 65}, 
      054028 (2002); {\bf D 66} 094009 (2002); {\bf D 66}; 014020 (2002); 
      {\bf D 67} 034008 (2003).
\bibitem{Chengsum} H-Y. Cheng, Int. J. Mod. Phys. {\bf A 21} 4209 (2006) 
      and references therein.
\bibitem{PL} M. Z. Wang {\em et. al.}, Belle Collaboration, 
             {\em Phys. Lett.} {\bf B 617}, 141 (2005).   
\bibitem{Harari} J. L. Rosner, Phys. Rev. Lett. {\bf 21}, 950 (1968); 
         {\bf 22}, 689 (1969); H. Harari, Phys. Rev. Lett. {\bf 22}, 562 (1969).
\bibitem{Ablikim}  M. Ablikim {\em et al.,} BES Collaboration, {\em Phys. Rev. Lett.} 
        {\bf 95}, 262001 (2005).    
\bibitem{JW} M. Jacob and G. Wick, {\em Ann. Phys.} {\bf 7}, 404 (1959).
\bibitem{Rose} M.E. Rose, {\em Elementary Theory of Angular Momentum},
        (Wiley and Sons, New York, 1957): 
        A.R. Edmonds, {\em Angular Momentum in Quantum Mechanics}
        (Princeton Univ. Press, Princeton, N.J., 1960).
\bibitem{Meissner} J. Haidenbauer, Ulf-G. Meissner, and A. Sibirtsev,
        {\em Phys. Rev.} {\bf D 74}, 017501 (2006).
\bibitem{Faddeev} L.D. Faddeev, {\em Soviet Phys. JETP,} {\bf 12}, 1014 (1961);
        {\em Soviet Phys. Doklody,} {\bf 6}, 384 (1961); {\bf 7}, 600 (1963).
\bibitem{nuclear} B. Kerbikov, A. Stavinsky, and V. Fedotov, {\em Phys. Rev.} 
        {\bf C 69} 055205 (2004); 
        B.S. Zou and H.C. Chiang, {\em Phys. Rev.} {\bf D 69}, 034004 (2004);
\bibitem{nuclear2} A. Sibirtsev, J. Haidenbauer, S. Krewald, Ulf-G. Meissner,
        and A. W. Thomas, {\em Phys. Rev.} {\bf D 71}, 054010 (2005). 
\bibitem{nuclear3}   B. Loisseu and S. Wycech, {\em Phys. Rev.} {\bf D 72}, 011001
        (2005).
\bibitem{Gillespie} J. Gillespie, {\em Final-State Interaction} (Holden
      Day, San Francisco, 1964), p.19 and references therein.
\bibitem{Jost} R. Jost and A. Pais, {\em Phys. Rev.} {\bf 82}, 840 (1951);
       R. Jost and W. Kohn, {\em Phys. Rev.} {\bf 87}, 977 (1952).  
\bibitem{Omnes} N. I. Muskhelishvili, {\em Singular Integral Equation}
      (Noordhoof, Gronigen, 1953), p.204; R. Omn{\' e}s, Nuovo Cimento
      {\bf 8}, 316 (1958).
\bibitem{Watson} K. Watson, {\em Phys. Rev.} {\bf 88}, 1163 (1952); K. Aizu, 
   {\em Proc. Intern. Conf. Theoret. Phys.}  (Kyoto-Tokyo Science Council, 
    1953), p.200; A. B. Migdal, {\em JETP}, {\bf 1}, 2 (1955).    
\bibitem{BES} S. Olsen, hep-ex/0305048; J.Z. Bai {\em et al.},
    BES Collaboration, {\em Phys. Rev. Lett.} {\bf 91}, 022001 (2003).   
\bibitem{gamma} K. Abe {\em et al}., Belle Collaboration,
          BELLE-CONF-0413, hep-ex/0409009.
\end{thebibliography}
\end{document}